\documentclass[10pt,letterpaper,twocolumn]{article} 

\usepackage{imaging}
\usepackage[draft]{hyperref}
\usepackage{amsmath}

\begin{document}

\twocolumn[ 

\title{Super-resolution single-beam imaging via compressive sampling}


\author{Wenlin Gong$^{*}$, and Shensheng Han$^{\dag}$}

\address{Key Laboratory for
Quantum Optics and Center for Cold Atom Physics of CAS, Shanghai Institute of Optics and Fine Mechanics, Chinese Academy of Sciences,
Shanghai 201800, China}
\address{$*$gongwl@siom.ac.cn}
\address{$\dag$sshan@mail.shcnc.ac.cn}

\begin{abstract}
Based on compressive sampling techniques and short exposure imaging, super-resolution imaging with thermal light is experimentally demonstrated exploiting the sparse prior property of images for standard conventional imaging system. Differences between super-resolution imaging demonstrated in this letter and super-resolution ghost imaging via compressive sampling (arXiv. Quant-ph/0911.4750v1 (2009)), and methods to further improve the imaging quality are also discussed.
\end{abstract}

 ] 

\noindent

Back in the 19th century, Lord Rayleigh had demonstrated that the best resolution of conventional imaging system was determined by the numerical aperture of the imaging lens \cite{Rayleigh}. Some decades later, the true limit on imaging arose from the optical wavelength $\lambda$ and the recoverable resolution was $\lambda/2$ exploiting the evanescent components of the electromagnetic field near the surface of object \cite{Ash,Betzig,Ramakrishna,Pendry}. Several digital image processing approaches can realize super-resolution, but they are highly sensitive to noise in the measured data and to the assumptions made on the prior knowledge \cite{Harris,Goodman,Hunt,DeGraaf}. In recent ten years or so, a new imaging technique called ghost imaging (quantum imaging) can realize subwavelength interference with the resolution beyond the classical diffraction limit by a factor of 2 using entangled source or thermal light source \cite{Angelo,Xiong,Angelo1}. When signals satisfied certain sparsity constraints, Donoho has demonstrated mathematically that super-resolution restoration is possible below the Rayleigh threshold ($\Omega \ll\frac{\pi}{\Delta}$, where $\Delta$ is the lattice span and $\Omega$ is the frequency cutoff)\cite{Donoho} and a new information processing techniques called as compressive sampling (CS) provides a solver for large-scale sparse reconstruction \cite{Candes1,Donoho1,Herman,Figueiredo,Candes2}. Combining the sparse prior property of images with the ghost imaging method, super-resolution ghost imaging via compressive sampling (GICS) with the resolution beyond the diffraction limit by a factor of 10 has been demonstrated experimentally even without looking at the object, and its physical principle suggests that using compressive sampling reconstruction algorithms, conventional imaging system may also realize super-resolution because all objects have sparse representations when expressed in the proper representation basis \cite{Gong}. For the conventional imaging system widely applied in people's life, as shown in Fig. 1, can we directly obtain super-resolution imaging with thermal light exploiting the sparse prior property of images and CS reconstruction algorithms?

\begin{figure}
\centerline{
\includegraphics[width=8.5cm]{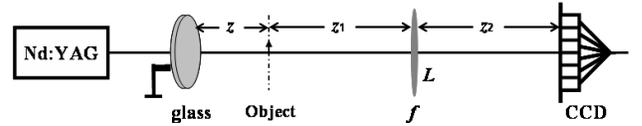}}
\caption{Standard schematic of conventional imaging and super-resolution imaging via compressive sampling with thermal light.}
\end{figure}

Here, we demonstrate super-resolution imaging via compressive sampling borne on thermal light using the standard conventional imaging system
shown in Fig. 1. The pseudo-thermal light source $S$, which is obtained by passing a laser beam (with the wavelength $\lambda$=532nm and the
source's transverse size $D$=4mm) through a slowly rotating ground glass disk, first goes through an object and then to an imaging lens with the focal length $f$. The intensity distributions transmitted through the object are finally recorded by a CCD camera. The distances $z_1$, $z_2$ and the focal length of the lens $f$ obey the Gaussian thin-lens equation: $\frac{1}{z_1}+\frac{1}{z_2}=\frac{1}{f}$. According to the Rayleigh criterion \cite{Rayleigh}, the resolution limit $\Delta x$ of conventional imaging is determined by the wavelength $\lambda$ and the numerical aperture (\emph{N.A.}) of the lens $f$, namely
\begin{eqnarray}
\Delta x=0.61\frac{\lambda}{N.A.}\simeq1.22\frac{\lambda z_1}{L}.
\end{eqnarray}
where $L$ is the effective transmission aperture of the imaging lens $f$.

Based on the theory of statistical optics \cite{Goodman1}, the intensity distribution on the detection plane at a certain time $s$
can be expressed as by the Fresnel diffraction integral \cite{Goodman}
\begin{eqnarray}
I_s (x, y)= \int {dx_0 }{dy_0 }{dx_0' }{dy_0' }E_s (x_0, y_0)E_s^*(x_0', y_0')\nonumber\\\times h^*(x, y; x_0', y_0')h(x, y; x_0, y_0).
\end{eqnarray}
where $E_s(x_0, y_0)$ denotes the light field on the source plane at time $s$, $h(x, y; x_0, y_0)$ and $h^*(x, y; x_0', y_0')$ are
the impulse function of optical system and its phase conjugate, respectively.

For the specific optical system depicted in Fig. 1 and under the paraxial approximation, the impulse response function of the optical system is
\begin{eqnarray}
h(x,y; x_0,y_0 ) \propto \int {dx'} {dy'} \exp \{ \frac{{j\pi }}{{\lambda z_1}}(x'^2+y'^2)\}\nonumber\\\times T(x',y')\exp \{ \frac{{j\pi }}{{\lambda z}}[(x' - x_0 )^2+(y' - y_0 )^2 ]\}\nonumber\\\times\sin c[\frac{L}{\lambda}(\frac{x}{z_2} + \frac{ x'}{z_1})]\sin c[\frac{L}{\lambda}(\frac{y}{z_2} + \frac{ y'}{z_1})]\}.
\end{eqnarray}
where $T(x,y)$ is the transmission function of the object, $\sin c(x)=\frac{\sin(\pi x)}{\pi x}$ and $\sin c(y)=\frac{\sin(\pi y)}{\pi y}$.

Similar to the idea of ghost imaging via compressive sampling (GICS) \cite{Gong}, for the schematic shown in Fig. 1, the CCD camera has
recorded the similar but not the same information among each observation because of the fluctuation of the light field, thus the intensity
distributions recorded with the CCD camera obey Gaussian statistical distribution and also satisfy the requirement of incoherence measurements
of CS. Different from GICS, both the sensing basis and the information transmitted through the object are recorded by the same CCD camera. If the
intensity distributions located at the CCD camera plane at time $s$ are described by $I_s(x,y)$, then the image of the object can be reconstructed
by solving the following convex optimization program \cite{Figueiredo}:
\begin{eqnarray}
\left| {T_{CS} } \right| = \left| {T'} \right|;{\rm{ \ }} {\rm{ subject \ to: }}{\rm{ \ }} \nonumber\\ \mathop {\min }\limits_{x,y}
{\rm{ \ }}\frac{{\rm{1}}}{{\rm{2}}}\left\| {\sum\limits_{x,y} {I_s (x,y)} } {- \sum\limits_{x,y}I_s (x,y)\left| {T'(x,y)} \right|^2 }
\right\|_2^2  \nonumber\\+ \tau \left\| {\left| {T'(x,y)} \right|^2 } \right\|_1 ,{\rm{ }}\forall _s  = 1 \cdots K.
\end{eqnarray}
where $K$ is the acquisition numbers, $\tau$ is a nonnegative parameter, $\left\| V \right\|_{2}$ denotes the Euclidean norm of $V$,
$\left\| V \right\|_{ 1 }  = \sum\nolimits_i {\left| {\upsilon_i } \right|}$ is the $\ell_1$ norm of $V$, and $\left| {T_{CS} } \right|$
is the object's function recovered by CS reconstruction algorithm. Moreover, $I_s (x,y)$ in Eq. (4) is
\begin{eqnarray}
I_s (x,y) \propto \int {dx_0 } {dy_0 } {dx_0' } {dy_0' } {dx} ' {dy'} {dx} '' {dy} ''  \nonumber\\\times E_s^ *  (x_0' ,y_0' )
E_s(x_0 ,y_0)T^ *(x'',y'')T(x',y')\nonumber\\\times\sin c[\frac{L}{\lambda}(\frac{x}{z_2} + \frac{ x'}{z_1})]
\sin c[\frac{L}{\lambda}(\frac{y}{z_2} + \frac{ y'}{z_1})]\nonumber\\\times\sin c[\frac{L}{\lambda}(\frac{x}{z_2}
+ \frac{ x''}{z_1})]\sin c[\frac{L}{\lambda}(\frac{y}{z_2} + \frac{ y''}{z_1})]\nonumber\\\times
\exp \{ \frac{{j\pi }}{{\lambda z}}[( x'-x_0)^2  - (x''-x_0')^2]\}\nonumber\\\times \exp \{ \frac{{j\pi }}{{\lambda z}}[( y'-y_0 )^2  - (y''-y_0')^2]\}
\nonumber\\\times \exp \{ \frac{{j\pi }}{{\lambda z_1 }}(x'^2  - x''^2 +y'^2  - y''^2 )\}.
\end{eqnarray}

In the experiment, the distances listed in Fig. 1 were as follows: $z$=200mm, $z_1$=$z_2$=800mm, and the focal length of the lens $f$=400mm. The effective transmission aperture of the imaging lens was $L$=2mm and the exposure time widow for the CCD camera was set to be 1 ms. The experimental results of imaging a double-slit and an aperture (``\textbf{zhong}" ring) are shown in Fig. 2. For comparison, Fig. 2(a) displays the images of the objects by an ordinary lens imaging system using thermal light. The images of the objects for conventional imaging, as depicted in Fig. 2(b), appear in the screen of the CCD camera. Using the method of imaging via compressive sampling and the gradient projection for sparse reconstruction (GPSR) algorithm \cite{Figueiredo}, we observe the superresolved images of the objects in Fig. 2(c).
\begin{figure}
\centerline{
\includegraphics[width=8.5cm]{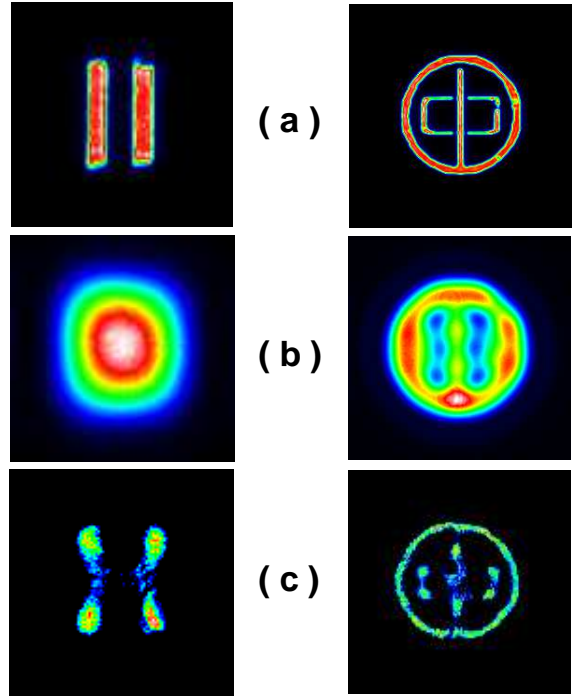}}
\caption{Experimental results of conventional imaging and super-resolution imaging via compressive sampling of two objects using the schematic
shown in Fig. 1. Left column: a double-slit (slit width $a$=90$\mu$m and center-to-center separation $d$=180$\mu$mm); right column:
an aperture (``\textbf{zhong}" ring). (a). The original object; (b) and (c) are the images reconstructed by conventional imaging and
super-resolution imaging via compressive sampling with thermal light, respectively (with 1000 observations).}
\end{figure}

The differences of super-resolution imaging demonstrated here and super-resolution ghost imaging shown in Ref. \cite{Gong} greatly deserve discussion. GICS is two-arm nonlocally imaging system and the diffraction limit of optical system is determined by the transverse coherence length on the object plane in the test path. In the image extraction process of CS reconstruction algorithm, the sensing basis in CS, highly correlated with the light field on the object plane in the test path, is registered by the CCD camera in the reference path and a bucket detector fixed in the test path has received all the information transmitted through the object. However, for the standard conventional imaging system shown in Fig. 1, because of the diffraction effect caused by the finite transverse size of the lens, the CCD camera records spatial low-resolution intensity distributions of the light field transmitting through the object and they are used to compose the sensing basis of CS reconstruction techniques. Apparently, the sensing basis of this two super-resolution imaging methods is different. For the former, the intensity distributions outside the field of the object also are incoherent among each observation and are helpful to the reconstruction of images. While for the latter, only the intensity distributions within the field of the object are useful to the measurements because the intensity distributions outside the field of the object are always zero for each observation (namely completely coherent). Therefore, the vector length of the representation basis for GICS will be much larger than that of conventional imaging system because it is the same as the vector length of sensing basis in CS reconstruction algorithms. Based on the characteristic of CS reconstruction theory, the larger the sparse degree of images in the representation basis is, the higher the probability of stable recovering the object will be \cite{Candes1,Donoho1,Herman}. Hence, GICS will have much higher ability to obtain superresolved images than that in conventional imaging system because the images of the same objects are much sparser in the representation basis with larger vector length. In addition, from the viewpoint of speckle, for the above two super-solution imaging methods with thermal light, super-resolution imaging relies on multiple short exposures and makes full use of the similar but not the same information among each observation. Based on the basic principle of speckle imaging \cite{Goodman1}, optical imaging measured over an exposure time on the order of several seconds (namely long exposure imaging) will give an average image and can not observe the intensity fluctuation of the light field. Completely different from long exposure imaging, short exposure imaging gives a speckled appearance and of course can catch the fluctuation of the light field. And the shorter the detection time is, the stronger the intensity fluctuation of the light field is and helpful to recovering the objects' images. Further, combining the improved images obtained by previous digital image processing approaches (such as deconvolution, analytical continuation) with suitable compressive sampling reconstruction algorithms, the ability of super-resolution may be further enhanced.

In conclusion, combining the sparse representation of the images in the proper representation basis with suitable compressive sampling
reconstruction algorithms, superresolved images can also be obtained in standard conventional optical imaging system. This imaging skill
can be directly applied to people's life and is very useful to the long-range imaging and microscopy.

The work is partly supported by National Natural Science Foundation of China under Grant Project No. 60877009,  National Natural
Science Foundation of China under Grant Project No. 60877009, and Shanghai Natural Science Foundation under Grant Project No. 09JC1415000.

\end{document}